\documentclass{article}
\usepackage[english]{babel}
\usepackage[utf8]{inputenc}
\usepackage{amsthm}
\usepackage{mathtools}
\usepackage{graphicx}
\usepackage[T1]{fontenc}
\usepackage{tabularx}
\usepackage{float}
\usepackage{amsfonts}
\usepackage{amsmath}
\usepackage{amssymb,latexsym}
\usepackage{color}

\begin{document}

\title{\bf  ASTRONOMY IN ISLAMIC WORLD: \\
A EUROPEAN PERSPECTIVE } 

\author{{\bf Richard Kerner}
\\ \\{\small  Laboratoire de Physique Th\'eorique de la Mati\`ere Condens\'ee} 
\\ {\small {\it Sorbonne-Universit\'e,}, {\it Paris, France}} \\  {\small e-mail: richard.kerner@sorbonne-universite.fr}}

%\hskip 0.2cm
%e-mail: richard.kerner@sorbonne-universite.fr

\date{ } % Leave empty to omit a date

\maketitle

\begin{abstract}
Mathematical and astronomical achievements of the Islamic World during its golden era are briefly exposed. 
Thie article is based on the invited talk delivered remotely at the ICRANET-Isfahan Astronomical meeting,
November 2-5, 2021, which, in turn, reproduces major parts of one of the chapters of my book ``Our Celestial
Clockwork'', published recently (2021) by the World Scientific. 
\end{abstract}

\section{The spread and the Golden Age of Islam} \label{sec:introduction}

%\title{\bf {\small ASTRONOMY IN ISLAMIC WORLD \\
%A EUROPEAN PERSPECTIVE }} 
%\author{{\bf Richard Kerner}
%\\{\it Laboratoire de Physique Th\'eorique de la Mati\`ere Condens\'ee} 
%\\ {\it Sorbonne-Universit\'e,}, {\it Paris, France}} %\date{30/06/2008 
%\date{{\color{violet}  {\bf ICRANet-Isfahan Astronomical Meeting }}\\
%{\color{violet} {\bf Isfahan University,}} \\
%{\color{magenta} {\bf Isfahan, IRAN}} \\
%{\color{blue} {\bf  November 3-5, 2021 }}} 
%
%\begin{document}
%
%\frame{\titlepage}
%
%\section{Summary} 
%
%\section{Introduction}
%
%
%\begin{figure}[hbt]
%\centering
%\includegraphics[width=8cm, height=4.8cm]{Old_World_820.png}
%\caption{\small{Islamic World in $820$ C.E.}}
%\label{fig:Islam620}
%\end{figure}

At the end of fifth century C.E. the Roman Empire was in shambles. It was divided into the Western and Easter parts
already in $395$ C.E., and prone to steadily increasing barbarian invasions ever since. The last Roman emperor Romulus Augustulus
was disposed of in $476$ C.E. by war Germanic chief Odoaker, who proclaimed himself emperor. 

%\begin{figure}[hbt]
%\centering
%\includegraphics[width=9cm, height=6.3cm]{Roman_Empire_Map.png}
%\caption{\small{Roman Empire at the height of its expansion, $180$ C.E. }}
%\label{fig:Rome180}
%\end{figure}

The Eastern part of the Empire with capital in Constantinople gradually transformed into Greek-speaking Byzantine Empire. All these territories, 
including Egypt and North Africa up to Spain were Christian since the Emperor Constantinus proclaimed Christianity to become state religion.  
However, since the fall of the Western Roman Empire, most of these territories were deprived of any organized 
army and statehood, prone to barbarian invasions, among which the Huns, the Goths and the Vandals were the most important and disastrous.

%\begin{figure}[hbt]
%\centering
%\includegraphics[width=9cm, height=6.4cm]{Rome_600AD.jpg}
%\caption{\small{Roman Empire after the fall, $580$ C.E. }}
%\label{fig:Rome580}
%\end{figure}
%
%\begin{figure}[hbt]
%\centering
%\includegraphics[width=9cm, height=5.7cm]{Persia_Rome7.png}
%\caption{\small{Eastern Roman Empire and Persia, $650$ C.E.} }
%\label{fig:Rome650}
%\end{figure}

%\section{The Golden Age}

In the midst of $7^{\rm th}$ century C.E. a new religion was proclaimed by Muhammad in Arabia,
derived from Jewish and Christian traditions, simplifying them and professing with great conviction the uniqueness of God, This was an important revolution, 
because it replaced the tribal principle by a more general religious and social unity.  

While Europe entered the era of turmoil and invasions Dark Ages, lasting roughly from $500$ C.E. 
 until the $12^{\rm th}$ century, the Arabs have conquered 
vast territories creating a buyoant Islamic Civilization, which spread from Moorish Spain in the West, through Western North Africa, 
through Egypt and Mesopotamia, touching India and even the most western parts of China. 

%\end{frame}%
%
%\begin{frame}{Islam in the $9^{\rm th}$ century C.E.}
%
%\begin{figure}[hbt]
%\centering
%\includegraphics[width=10cm, height=6.3cm]{spread.jpg}
%\caption{\small{The spread of Islam from $661$ till $1258$ C.E.}}
%\label{fig:Islam620}
%\end{figure}
The main stages of this extraordinary politically-religious adventure can be represented as follows:
\vskip 0.2cm
\indent 
$\bullet$ \; \; Rightly Guided Caliphs 632-661
\vskip 0.1cm
\indent
$\bullet$ \; \; 
Umayyad expansion 661-750
\vskip 0.1cm
\indent
$\bullet$ \; \; 
Abbasid expansion 750-1258
\vskip 0.1cm
\indent
$\bullet$ \; \; 
Moorish Spain 756 - 1162
\vskip 0.2cm
The only precedent of as rapid
expansion was the Hellenistic empire built by Alexander the Great. Like ancient Greek, and Latin in the Roman Empire, Arabic
language became universal among Islamic scholars, with Persian and Turkish as subsidiary in the East.  
Ancient Greek science has been assimilated and developed further. Aristotle, Euclid and Ptolemy
were translated and multiplied in many copies.
Medical science of ancient Greeks was also honored, and Hippocrates and Galen were translated, too. 

The {\it Golden Age} of Islamic civilization covers the time span from $9^{\rm th}$ till $13^{\rm th}$ century C.E. in Baghdad under
the {\it Abassid} dynasty, and in Moorish Spain until the fourteenth century; it lasted a little bit more in Central Asia, 
falling under the Mongol conquest. 

During the first 115 years of the Abbasid Caliphate, (from 750 to 1258 CE) the Islamic State saw 
a record growth in the fields of the arts of literature and music, the sciences (especially astronomy), philosophy, mathematics, medicine, 
culture, commerce, and industry. Arabic, the language of the Quran became the language of international scholarship. In addition to the capital, 
Baghdad, many provincial centers competed with each other. A number of scientific terms of arabic origin are in use in Western science
until today, and known to everybody: it is enough to cite a few, like {\it algebra}, {\it algorithm}, {\it zenith}, {\it azimuth},
{\it nadir}, {\it alcohol}, {\it alcali}. 

In $751$ C.E., with the help of a few Chinese prisoners, the Abbasid governor of Samarkand founded the paper industry. 
In $800$ C.E. paper mills were established in Baghdad and Damascus; in $900$ C.E. another one was established in Cairo. By $1150$ C.E. several 
were established in Morocco and Spain. As a result, Islamic learning spread rapidly into Europe.

In 756 CE, Amir Abd ar-Rahman came to power in Spain. His term also contributed to the Islamic Golden Age. He organized a system of law and justice 
and was very fond of knowledge and learning. 
Scholars from all over Europe sought the knowledge and learning from Spain during his term. Islamic Universities were the only educational institutions 
free of charge.

In 830 CE the seventh Abbasid Caliph, al-Mamun, established the famous House of Wisdom in Baghdad. The Greek language gave way to Arabic 
as a form of expression of scientific and philosophical ideas. Classical Greek literature was translated into Arabic and Arabic speaking scholars 
wrote a number of renowned commentaries.

No wonder that during the awakening of Medieval Western Europe, the scholars first turned to Arabic translations of ancient Greek science.
In order to enrich Christian Europe in scientific, medical and philosophical studies, these works had to be retranslated into Latin. 
This was mainly done in Muslim Spain and Sicily and these books served as textbooks in universities for centuries.

However; recognizing the crucial role of Islamic Golden Age in in the transmission of ancient Greek science, we should not forget the fact
that it was previously conserved and transmitted by Romans and Greeks themselves, especially during the early Christian era. The first
Arabic translations were made no earlier than in the midst of the $8^{\rm th}$ century, $1100$ years after Aristotle, $1050$ nyears after
Euclid and $900$ years after Ptolemy published their scientific paperrs in Greek, which were written originally on a relatively fragile
papyrus support. An average papyrus, especially if it is used for frequent reading, would last no more than a couple of centuries; this
means that the papyri from which the Arabs translated the works of Aristotle, Euclid and Ptolemy were already copied many times and translated
into Syriac by Christians of the Eastern Roman Empire. Before that, Latin translations were made in the West and conserved in some remote
monasteries, mostly in Ireland, and forgotten. The bnew impulse came from Baghdad and Spain during the Golden Age of Islam.  

\section{The sources}

Islamic science took its sources on both sides of its geographical spread, namely in Greek scientific tradition conserved
in Byzantium and in Hellenistic Egypt, both Christianized since many centuries, and in Indian and Persian mathematical and astronomical traditions. 

The Greeks owned much of astronomical knowledge to the earlier Babylonian and Egyptian civilizations; however, they enriched
and extended them in unprecedented proportions.

In ancient Babylon, the observations were performed by Chaldeians from the top of pyramidal constructions called {\it ziggurats}.
They were able to determine with astonishing precision the celestial coordinates of stars and planets, and follow the Sun's annual and the Moon's 
monthly motions. As in the ancient Egypt, timekeeping was one of their most important functions, as well as interpreting omens from planetary 
configurations and other noticeable celestial events, especially solar and lunar eclipses, some of which they came
to predict quite successfully due to the comprehension of the so-called {\it saros} periodicity. 

The earliest acquaintance of Islamic world with mathematics and astronomy was due to Indian and Persian influences. From
India, the Arabs took the decimal system with numerals and zero which we use today. The main Indian mathematical text by
Brahmagupta ($598 - 670$ C.E.) was very early translated into Arabic. 

It contained, among others, the idea of negative numbers and their algebraical properties. They were called ``debts'', in opposition 
to positive numbers called ``fortunes''. Not only their algebraical
addition was defined, but also the multiplication, with well-known rules, including the fact that a product of two negative
numbers is a positive one. 

Brahmagupta owed much of his knowledge to his predecessor Aryabhata ($476-550$ C.E.), duly called the ``father of Indian mathematics''.
Aryabhata was undoubtedly the greatest mathematician and astronomer of ancient India. 
His major work is known as {\it Aryabhatiya}. It contained the elements of spherical trigonometry, quadratic equations, algebra, 
plane trigonometry, sums of power series and arithmetics.

%\end{document}
%
%\begin{figure}[hbt]
%\centering
%\includegraphics[width=2.8cm, height=3.2cm]{Aryabhatta_color.jpg}
%\hskip 0.4cm
%\includegraphics[width=3.1cm, height=4.5cm]{Aryabhata.jpg}
%\caption{\small{\color{violet} {\bf {\color{black} Aryabhata} {\color{black} $476-550$ C.E.} {\color{blue} Aryabhata } is the father of Indian mathematics. 
%He was the greatest mathematician and astronomer of ancient India. 
%His major work is known as {\color{red} Aryabhatiya}. It consists of {\color{blue} spherical trigonometry, quadratic equations, algebra, 
%plane trigonometry, sums of power series, arithmetic}.}}}
%\label{fig:Aryabhata}
%\end{figure}

On the western side of the Arabs' early advance were the vast territories of the christianized Roman Empire, heir to Latin and Greek culture and
civilization. The three pillars of Greek science which were adopted by Arab and Islamic polymaths were: Aristotle for physics 
and all other natural sciences, Euclid for geometry, and Ptolemy for astronomy. Their major works were translated, copied, 
edited and commented upon by Muslim scholars for centuries to come. 

Aristotle would assume the title of ``First Teacher'', vouching for all the intellectual heritage of the Greeks, viewed as a systematic 
knowledge compatible with the Muslim worldview. 
Together with Aristotle's own works, the Arabs translated an equally great amount of commentaries tracing back to late Antiquity, 
particularly to the School of Alexandria. Some of Plato's texts were translated at the same time.

%\begin{figure}[hbt]
%\centering
%\includegraphics[width=3.3cm, height=4.6cm]{Plato_Aristotle.jpg}
%\caption{\small{\color{violet} {\bf  Plato and Aristotle, as they appear on Rafael Santi's monumental painting ``The School of Athens''}}}
%\label{fig:Rafael}
%\end{figure}
%
%\begin{figure}[hbt]
%\centering
%\includegraphics[width=3.1cm, height=4.2cm]{Aristote_Louvre.jpg}
%\hskip 0.5cm
%\includegraphics[width=3.1cm, height=4.4cm]{Aristotle_Hebrew.jpg}
%\caption{\small{\color{violet} {\bf Left: {\color{red} Aristotle's} portrait in the Louvre; 
%Right: A page from Aristotle's {\color{blue} ``Organon''} in Hebrew translation }}}
%\label{fig:Aristotle}
%\end{figure}
%
%
%\begin{figure}[hbt]
%\centering
%\includegraphics[width=4cm, height=3.2cm]{igeotools.jpg}
%\hskip 0.3cm
%\includegraphics[width=2.2cm, height=3.3cm]{Euclidgrec.jpg}
%\caption{\small{\color{violet} {\bf Left: Tools of Euclidean geometry, Right: Euclid, {\color{black} $323-265$ B.C.E.} 
Another great contribution to the development of science in the Islamic world came from ancient Greece via the translation of the
totality of books written by Euclid. The great Greek mathematician Euclid united in the $13$ books of his major work,
 ``The Elements'' the geometrical knowledge elaborated by his predecessors, Pythagoras ($570 - 495$ B.C.E.), 
Hippocrates of Chios ($470-410$ B.C.E), Eudoxus of Cnidos ($406-355$ B.C.E.). Euclid's merit was not only the systematization of all 
geometrical facts and theorems known in his time, but above all, the proof that they can be derived from only FIVE POSTULATES.

%\begin{figure}[hbt]
%\centering
%\includegraphics[width=3.7cm, height=3cm]{Euclidpap.jpg}
%\hskip 0.3cm
%\includegraphics[width=2.8cm, height=4cm]{EuclidStatueOxford.jpg}
%\caption{\small{\color{violet} {\bf Left: One of the rare fragments of ``Elements'', Right: Euclid's statue in Oxford.

The ``Elements'' were copied many times in Greek in Byzantium, and later on translated in Syriac. The Arabic translation
from Syriac and Greek appeared around $800$ C.E. under the Khalif Haroun al-Rachid. In $1120$ 
English monk Adelard of Bath translated Euclid's book from Arabic into Latin. 

However, ancient Greek mathematics was mostly geometry, which was conceived as more basic than arithmetics. In fact, Greeks 
made a clear distinction between ``numbers'', which were reduced to integers, fractions and rational numbers, and ``magnitudes''
such as the diagonal of a square, which was proven to be not expressible as ratonal number (fraction). Greek calculus
was marred by the absence of adequate notation system including the extremely useful concept of zero, present in Indian
mathematics. The Arabs were able to merge the two notions into a single one, including negative numbers and incommensurable
quantities like square roots.  

Astronomy, which in these times was synonymous with astrology, was given great importance by Muslim rulers since the advent of Islam.
The main bulk of astronomical knowledge was contained in Ptolemy's ``Magna Opus'', translated into Arabic from Syriac and Greek
versions conserved in Christian libraries that survived the fall of Roman Empire.  

Claudius Ptolemy lived in the Egyptian city of Alexandria from $100$ till $170$ C.E. or even till $178$ C.E. according to
some Arab commentators and translators of his writings. At that time Egypt was a Roman province
ruled by the Greek dynasty bearing the same family name, {\it Ptolemaios}.

Ptolemy was the last great representative of Greek science, 
heir to a tradition eight centuries old. Ptolemy lived and worked $400$ years after Euclid, $600$ years after Pericles
and $700$ years after Anaximander of Miletus who was the first to suggest the idea of the celestial sphere and to produce a map 
of the inhabited world known in his time. 

Little is known about Ptolemy's life except that he probably spent it all in Alexandria, where he observed two solar eclipses: one
in the ninth year of Hadrian's rule ($125$ C.E.) and in the fourth year of Antoninus' rule ($141$ C.E.). Besides his {\it Syntaxis Mathematicus},
known also as ``Magnum Opus" due to its importance for astronomy, and mostly under the name of its arabic translation 
{\it Almagest}. Ptolemy was also a renowned geographer in his time.         

The geocentric system elaborated in {\it Syntaxis} was universally accepted during Hellenic
and Roman antiquity, and later on by medieval Christian and Muslim civilizations alike, and was used to describe and predict motions
of celestial bodies by astronomers and astrologists.

\section{Islamic Science and Scholars}

Mathematics and astronomy in Islamic world developed in a larger context of cultural, philosophical, religious and scientific
development, due to remarkable thinkers, physicians and polymaths, among whom the greatest ones were the following: 
\vskip 0.2cm
\indent
$\bullet$ \; {\bf Al-Farabi}, ($..? - 950$), known in Christian world under the latinized name {\it Alpharabius}. 
He spent most of his life in Baghdad. A prolific
writer, he authored comments on Platonic and Aristotelian philosophy,  and original works on physics, alchemy, psychology, astronomy and music. 
His authority was so great in Islamic world, that he was often called ``The Second Teacher" - meaning second after Aristotle.
His comments paved the way to the philosophical synthesis by Ibn Sina (Avicenna) and Ibn Rushd (Averroes). 
In his late years he moved to Damascus,  where he died in $950$ C.E.
\vskip 0.1cm
\indent
$\bullet$ \; {\bf Avicenna (Ibn Sina)}, ($.;? - 1037$),
renowned physician and philosopher, called ``The Third Teacher''. His most appreciated work was the {\it Canon of Medicine},
a manual of medical science in many volumes, which became the authority for European doctors after it was translated into Latin. 
Among his lasting contributions was the drug testing method and the discovery of the infectiousness of tuberculosis.

%\begin{figure}[hbt]
%\centering
%\includegraphics[width=9cm, height=5.6cm]{Avicenna_Persian_Galen.jpg}
%\caption{\small{\color{violet} {\bf {\color{red}  Avicenna} (Ibn Sina, $..? - 1037$, often referred to as {\color{blue} ``the Persian Galen''}.}}}
%\label{fig:Persian_Galen}
%\end{figure}
%
%
%\begin{figure}[hbt]
%\centering
%\includegraphics[width=9cm, height=5cm]{Avicenna_Isfahan.jpg}
%\caption{\small{\bf {\color{red} Avicenna (Ibn Sina)} {\color{violet}  at the court of the Governor of Isfahan}.}}
%\label{fig:AvicennaIsfahan}
%\end{figure}
%
%
%\begin{figure}[hbt]
%\centering
%\includegraphics[width=9cm, height=5cm]{Avicenna_Odyssey.jpg}
%\caption{\small{\bf {\color{red} Avicenna's (Ibn Sina) } {\color{violet} peregrinations during his life. }}}
%\label{fig:AvicennaOdyssey}
%\end{figure}

%\vskip 0.1cm
%\indent
%$\bullet$ \;  Al-Biruni, ($..? - 1048$), a famous geographer and astronomer, has improved Eratosthenes' estimate of Earth's circumference. 
%He also wrote the geographic description of India, a source of new historical, botanical and zoological facts. 
%Unfortunately, never translated into Latin before the modern times.  

\vskip 0.1cm
\indent
$\bullet$ \;  {\bf Averroes (Ibn Rushd)}, ($1126 - 1198$). Considered as one of the greatest Muslim scientists, he was born in Andalusia, then ruled by
the Almoravid dynasty, later passed under the Almohads. He studied diligently the writings of Aristotle, Al-Farabi 
and Ibn-Sina (Avicenna). He became the most influential Islamic scientist among the Europeans; his writings were studied in Padova, Salamanca, Oxford 
and Sorbonne. 

In $1189$ the Almohad caliph al-Mansour prohibits practicing music, poetry, and overall, philosophy. Averroes is accused of heresy and banished 
first from Cordoba, then from Andalusia to Morocco; his books are burned and prohibited. Although partly pardoned, he died in Morocco, never
being permitted to come back to his hometown Cordoba. 

\vskip 0.1cm
\indent
$\bullet$ \; {\bf Omar Khayyam } ($1048-1131$), a renowned mathematician, astronomer, philosopher and poet. Born in Nishapur, of Persian origin, he received 
an excellent education in his native Nishapur, and afterwards moved to Samarkand (now in Uzbekistan). His mathematical reputation rests principally
on his {\it ``Treatise on Demonstration of Problems of Algebra''}. He made such a name for himself that the Seljuq sultan Malik-Shah invited 
him to Isfahan to undertake the astronomical observations necessary for 
the reform of the calendar. To accomplish this an observatory was built there, and a new calendar was produced, known as the Jalali calendar. 
Based on making 8 of every 33 years leap years, it was more accurate than the present Gregorian calendar, and it was adopted in 1075 by Malik-Shah.

%
% In this treatise he gave a systematic discussion of the solution of cubic equations by means of intersecting conic sections. 
%Perhaps it was in the context of this work that he discovered how to extend Abu al-Wafa's results on the extraction of cube and fourth roots 
%to the extraction of n-th roots of numbers for arbitrary whole numbers n.
%
%He made such a name for himself that the Seljuq sultan Malik-Shah invited him to Isfahan to undertake the astronomical observations necessary for 
%the reform of the calendar. To accomplish this an observatory was built there, and a new calendar was produced, known as the Jalali calendar. 
%Based on making 8 of every 33 years leap years, it was more accurate than the present Gregorian calendar, and it was adopted in 1075 by Malik-Shah. 
%In Isfahan he also produced fundamental critiques of Euclid's theory of parallels as well as his theory of proportion. In connection with the former 
%his ideas eventually made their way to Europe, where they influenced the English mathematician John Wallis (1616-1703); in connection with the latter 
%he argued for the important idea of enlarging the notion of number to include ratios of magnitudes (and hence such irrational numbers 
%as $\sqrt{2}$ and $\pi$).

His years in Isfahan were very productive ones, but after the death of his patron in $1092$ he fell in disgrace, and returned to Nishapur 
where he taught and served the court as an astrologer. 
%Philosophy, jurisprudence, history, mathematics, medicine, and astronomy are among the subjects mastered by this brilliant man.

Omar's fame in the West rests upon the collection of robaiyat, or quatrains, attributed to him. 
 
%These quatrains have been translated into almost every major language and are largely responsible 
%for colouring European ideas about Persian poetry. Some scholars have doubted that Omar wrote poetry. His contemporaries 
%took no notice of his verse, and not until two centuries after his death did a few quatrains appear under his name. Even then,
% the verses were mostly used as quotations against particular views ostensibly held by Omar, leading some scholars to suspect that 
%they may have been invented and attributed to Omar because of his scholarly reputation.

As Bernard Lewis ($1916-2018$) has put it in his book {\it What went wrong?}, for many centuries the world of Islam was in the forefront 
of human civilization and achievement. However, due to many interior and exterior factors, Islamic science stopped its development around 
fifteenth century; since then, European science took the lead.

%\begin{figure}[hbt]
%\centering
%\includegraphics[width=9cm, height=5cm]{Averroes_Cordoba.jpg}
%\caption{\small{\color{violet} {\bf Averroes in Cordoba.}}}
%\label{fig:Averroes}
%\end{figure}

\section{Mathematics}

One of the most often repeated claims is that Muslims invented algebra. This is largely true, even if initially they took source 
in ancient Greek and Indian mathematics. The word ``algebra" comes from the arabic {\it al-jabr} and means ``restoring" or ``putting parts together". 

It was coined by Persian mathematician and astronomer al-Khwarizmi (ca. 780 - ca. 850 C.E.) who lived in Baghdad, in his treatise 
{\it ``llm al-jabr wa al-muqabala"}, which means ``Science of putting parts together and balancing''.  
In modern algebraical terms, the first word referred to the possibility of replacing terms from one side of an equation to another, 
and adding or substracting the same quantity from both sides. Al-Kwarizmi's name refers to the city of Khorezm (today in Uzbekistan) and was known 
in the West in its latinized version {\it Alkorizm} or {\it Algorithm}, and became the name of a general prescription in mathematics. 

%\begin{figure}[hbt]
%\centering
%\includegraphics[width=6cm, height=3.8cm]{AlKhorezmi_sculpture.jpg}
%\hskip 0.3cm
%\includegraphics[width=2.7cm, height=3.8cm]{AlKhorezmi_timbre.jpg}
%\caption{\small{\color{violet} {\bf Left: Al-Khorezmi's statue, Right: Postal stamp with Al-Khorezmi, USSR,  }}}
%\label{fig:Khorezmi}
%\end{figure}

Al Khwarizmi's book contained plenty of examples how to solve problems involving commerce, trade, inheritance, marriage 
and redemption of slaves. The examples did not involve any algebraic symbols yet, using geometrical figures modelling
the relations between numbers. But in other Arabic mathematical treatises symbols, numbers and words tended to replace
geometrical constructions. This was a major revolution in mathematics, with enormous impact on its further development.

For the mathematicians of ancient Greece a major difference existed between {\it numbers} and {\it magnitudes}, the second ones meaning
measurable lengths, areas or volumes. Numbers were used as long as they could be expressed as integers or their ratios, e.e. 
fractions. These quantities were called ``rational numbers'', in contrast with ``irrational ones'', which could be constructed geometrically,
but not algebraically, like the most famous square root of $2$, which is simply the diagonal of a unit square, but no finite fraction
exists whose square would be equal to $2$. 

The new approach to mathematics made possible simple and elegant proofs in place of sophisticated geometric constructions. For example,
whereas the well known geometrical theorem saying that the area of a square with sides $a+b$ is equal to the area of a square with side $a$
plus the area of another square with side $b$ plus the area of two rectangles of sides $a$ and $b$ can be proved with no effort by an obvious
geometric construction, the proof of another formula, that the area of a rectangle with sides $(a+b)$ and $(a-b)$ (with $a > b$) is equal to the  
difference between the areas of squares with side $a$ and with side $b$, i.e. $(a+b) \cdot (a-b) = a^2 - b^2,$ needs a more sophisticated 
geometrical consgruction,
%as shown in the Figure (\ref{fig:absquare}) below:
%\begin{figure}[hbt]
%\centering
%\includegraphics[width=3.6cm, height=3.6cm]{(a+b)2.jpg}
%\hskip 0.2cm
%\includegraphics[width=6.2cm, height=3.4cm]{a2-b2_B.png}
%\caption{{\small Ancient Greeks' derivation of algebraic formulas $(a+b)^2 = a^2 + 2 ab + b^2$ (left) and $(a+b)\cdot (a-b) = a(a+b)-ab-b^2 = 
%a^2 - b^2$ (right). }}
%\label{fig:absquare}
%\end{figure}
Among the shortcomings of the purely geometrical approach of the ancient Greeks was the lack of interest in the possibility of negative numbers
and arithmetical operations involving negative entities, which were known in ancient China since $300$ B.C.E. Islamic mathematicians not
only incorporated negative numers in algebraic operations, but also started to treat rational and irrational numbers on the same footing,
making no distinction when adding, substraction of multiplying them. 

The advantage of the algebraic approach is that it can be easily generalized beyond its usefullness in describing geometrical constructions
in two and three dimensions. The formula $(a+b)^2 = a^2 + 2 ab + b^2$ can be generalized to three dimensions geometrically, giving the volume of
a cube with side $(a+b)$ as a sum of one cube with side $a$, another cube with side $b$, three parallelepipeds with base $a^2$ and length
$b$, and three parallelepipeds with base $b^2$ and height $a$, as follows: 
$(a+b)^3 = a^3 + b^3 + 3 a^2 b + 3 a b^2.$
But nothing forbids to write down similar formulas for any power of the sum $(a+b)$, e.g. 
$(a+b)^4 = a^4 + 4 a^3 b + 10 a^2 b^2 + 4 a b^3 + b^4,$ and so on.  
       
Nevertheless most of Muslim authors did not consider negative numbers on the same footing as the positive ones. They knew how to solve linear
 and quadratic equations, but whenever there was a negative solution, they rejected it as ``absurd". 

An important step forward due to Muslim mathematicians is the establishment of new  arithmetic algorithms:  the extension of root-extraction 
procedures, known to Hindus and Greeks only for square and cube roots, to roots of higher degree and by the extension of the Hindu decimal 
system for whole numbers to include decimal fractions as computational devices. Omar Khayyam ($1148-1131$ C.E.), 
Persian mathematician, philosopher,
astronomer and  poet (remembered more for his beautiful poetry than for mathematical achievements) considered the general problem of extracting 
roots of any desired degree. He also closely approached the general solution of cubic equations, and elaborated a unifying approach to geometry
 and algebra.

Islamic algebraists of the $10^{\rm th}$ century made a substantial progress, generalizing al-Khwarizmi's quadratic polynomials to the algebra 
of expressions 
involving arbitrary positive or negative integral powers of the unknown. Several algebraists explicitly stressed the analogy between the rules 
for working with powers of the unknown in algebra and those for working with powers of $10$ in arithmetic. 

Similar progress was made in geometry. The Islamic mathematicians Thabit ibn Qurrah ($836-901$), his grandson Ibrahim ibn Sinan ($909-946$), 
Abu Sahl al-Kuhi (died c. $995$), and Ibn al-Haytham ($965-1040$), known in Europe as Alhazen, solved problems involving the geometry of conic sections, 
including the calculus of areas and volumes of plane and solid figures formed from them. They also investigated the optical properties of mirrors 
made from conic sections, which became of crucial importance centuries later, for the construction of optical devices.

\section{Astronomy}

During the ``Golden Age'' of Islam, astronomy was one of the most important sciences. Muslim religion had imposed several requirements
which needed important astronomical skills to improve the time-keeping and space orientation. Exact timing of five obligatory
prayers a day, beginning and the end of fast during the holy month of Ramadan, linked to the first and last crescent of the Moon, establishing 
times of islamic holidays - all that created the need for astronomical observations. 

Newly constructed Mosques had to be oriented towards Mecca in the Arabian
 Peninsula, and so should be positioned a Muslim during his prayer - this alone needed improved geographical and astronomical knowledge, too.

 Finally, vast territories of Islam, with flourishing terrestrial and maritime trade, made necessary the emergence of practical astronomy,
 improving observational devices like sextants and astrolabiums, to be used on land or in the open sea. 

Astronomy became the chief amongst the sciences in which Muslim scholars excelled. Success was built upon mastery of mathematics - geometry in particular 
- as much as the philosophical inheritance of the ancient Greeks and other early civilizations through translation into Arabic. 
Observatories founded in the most important centres of learning such as Baghdad and Cairo required specialized instruments for observation and instruction. 
These included celestial globes, quadrants, and armillary spheres, but the most sophisticated instrument adopted and developed by Muslim astronomers 
was the {\it astrolabe}. 

%\begin{figure}[hbt]
%\centering
%\includegraphics[width=4cm, height=4cm]{Astrolabio_Toledo1067.jpg}
%\hskip 0.2cm
%\includegraphics[width=4cm, height=5.5cm]{Armillary_sphere.jpg}
%\caption{\small{ Left: Ancient Astrolabe, Toledo $1067$ C.E.; 
%Right: Armillary sphere from Damascus, $1120$ }}
%\label{fig:Astrolabio}
%\end{figure}

How great was the impact of the Arabs on European astronomy that took the lead after $16^{\rm th}$ century 
can be easily seen in any modern stellar atlas. Here are a few among the best known bright stars whose names are of the Arab origin:
\vskip 0.2cm
\indent
Aldebaran, $\alpha$ Tauri, from arabic {\it Al-Dabaran}, the ``Follower";
\vskip 0.1cm
\indent
Algol, $\beta$ Persei, from arabic {\it Al Ghoul}, the ``Demon", or ``Scary monster";
(Named so probably because its periodical blinking in matter of hours could be easily seen by a naked eye). 
\vskip 0.1cm
\indent
Altair, $\alpha$ Aquilae, from arabic {\it an-Nisr ut-Ta'ir}, the ``Flying Eagle";
\vskip 0.1cm
\indent
Betelgeuse, $\alpha$ Orionis, from arabic {\it Yad al Jawza}, the ``Hand of Al-Jawza", a mythical character;
\vskip 0.1cm
\indent
Deneb, $\alpha$ Cygni, from arabic {\it Dhaneb ud-Djadjab}, ``Hen's tail"; 
\vskip 0.1cm
\indent
Dubhe, $\alpha$ Ursae Majoris, from arabic {\it Dubb}, ``Bear";
\vskip 0.1cm
\indent
Fomalhaut, $\alpha$ Piscis Austrini, from arabic {\it Fum al-Hul}, the ``Mouth of the Whale".

%\begin{figure}[hbt]
%\centering
%\includegraphics[width=7.8cm, height=4.8cm]{Muslim_Astronomers.jpg}
%\caption{\small{\color{violet} {\bf Muslim Astronomers working around {\color{red} Taqi-ad-Din} at the 
%{\color{blue} Istanbul Observatory}, from and old Turkish book.}}}
%\label{fig:Astronomers}
%\end{figure}

The Arab astronomers continued observations using the techniques inherited from Greeks and incorporating other ancient traditions, e.g. from
Persia and India. They were often able to improve and enrich the data collected by Ptolemy in his {\it Almagest}. An example: 
following Hipparchus, Ptolemy estimated precession of the equinoxes to be of $1^o$ per $100$ years, which would
correspond to completing the full cycle in $36 000$ years. Egyptian astronomer Ibn Yunus ($850 - 1009$ C.E.) has corrected 
this estimate, reducing it down to $1^o$ in $70$ years, which corresponds to the full cycle of $25 200$ years, very close to the presently accepted value. 
Let us present the most remarkable astronomers and their contributions.

%{\small{\color{blue} {\bf  The most eminent Muslim astronomers include Al-Battani, Al-Sufi, Al-Biruni, Al-Bitruji, Ibn Yunus, al Tusi and al Shatir. }}}

\vskip 0.2cm
\indent
$\bullet$ \; {\bf  Al-Khwarizmi} (ca. $786-850$ C.E.), 
of Persian origin, was appointed by al-Mamoun, the son of Harun-al-Rashid of the Abbasid dynasty, as the chief astronomer and head of the library in the 
``House of Wisdom'' in Baghdad. 

Al-Khwarizmi's important strides in astronomy and geography include the measurement of the length of a degree of a meridian in the plain of Sinjar,
improving Erastothenes' evaluation of Earth's circumference, the creation of a world map based on the geography of Ptolemy,
providing more accurate coordinates of approximately $2 \;400$ sites in the known world. These results were contained in his book
``Kitab surat al-ard'' (``The Image of the Earth'') 

Another al-Khwarizmi's treaty that made it into the Western canon of mathematical studies was a compilation of astronomical tables 
including a table of sines, and  was translated into Latin. He also produced two treatises on the sundial, on the astrolabe, 
and one on the Jewish calendar. Around $1110$ the Englishman Robert of Chester, who was then in Spain, translated into Latin 
Al-Khwarizmi's al-Jabr (Algebra) and introduced into Europe Arabic numerals, 
including the figure zero, which the Arabs themselves had borrowed from the Indian mathematicians.
\vskip 0.1cm
\indent
$\bullet$ \;  {\bf Al-Battani} (died in 929), known to Europe as Albategni or Albatenius was the author of the {\it Sabian tables} (al-Zij al-Sabi), 
a work which had great impact on his Muslim and Christian successors. His improved tables of the orbits of the sun and the moon comprise 
his discovery that the direction of the sun's eccentric as recorded by Ptolemy was changing. This, in modern astronomy, means the Earth's 
line of apsides is slowly moving. He also worked on the timing of the new moons, the length of the solar and sideral year, the prediction of eclipses, 
and the phenomenon of parallax. 

%
%\begin{figure}[hbt]
%\centering
%\includegraphics[width=6.8cm, height=4.5cm]{Al-Battani_Tables.jpg}
%\caption{\small{\color{violet} {\bf Al-Battani, called by the contemporaries ``The Second Ptolemy'' }}}
%\label{fig:AlBattani}
%\end{figure}

Al-Battani was also a pioneer in the field of trigonometry. He was among the first, if not the first, to use trigonometric ratios 
as we know them today. During the same period, Yahya Ibn Abi Mansour had completely revised the Zij of Almagest 
after meticulous observations and tests producing the famous Al-Zij al Mumtahan (the validated Zij). In al-Zij al-Sabi (known as the Sabian Tables), 
Al-Battani plotted the Sun's orbit more accurately than Ptolemy did. 
In his {\it ``De Revolutionibus''}, Copernicus refers to Al-Battani's calculations no fewer than 23 times. 
\vskip 0.1cm
\indent
$\bullet$ \; Belonging to the same era, {\bf Abd-al Rahman al-Sufi} ($903-986$) made several observations on the obliquity of the ecliptic 
and the motion of the sun (or the length of the solar year). He became renowned for his observations and descriptions of the stars, their positions, 
their brightness and their colour, setting out his results constellation by constellation. For each constellation, he provided two drawings, 
one from the outside of a celestial globe, and the other from the inside (as seen from the sky). Al-Sufi also wrote comments on the astrolabe, 
finding numerous additional uses for it (including one's location, measuring distances and heights, etc.) 
\vskip 0.1cm
\indent
$\bullet$ \; Egyptian astronomer {\bf Ibn Yunus} (died in 1009), in his observation endeavours included, amongst others more than 10, 000 entries 
of the Sun's position throughout the years using a large astrolabe of nearly 1.4 m in diameter. His work, in French edition, 
was centuries later an inspiration for Laplace in his determination of the ``Obliquity of the Ecliptic'' and the ``Inequalities of Jupiter and Saturn''. 

The famous European astronomer Newcomb also used his observations of eclipses in the motions of the moon. 

%
%\begin{figure}[hbt]
%\centering
%\includegraphics[width=3.8cm, height=4.7cm]{IbnYunus_al-Masri.jpg}
%\hskip 0.2cm
%\includegraphics[width=3.6cm, height=5cm]{Ibn_Yunus_Zij.jpg}
%\caption{\small{\color{violet} {\bf  Left: Ibn Yunus al Masri, Right: A page from Ibn Yunus book al-Zij }}}
%\label{fig:Ibn_Yunus}
%\end{figure}
%
\vskip 0.1cm
\indent
$\bullet$ \; {\bf Al-Biruni} ($973-1050$) claimed that the Earth rotated around its own axis. He calculated Earth's circumference, and fixed scientifically 
the direction of Mecca from any point of the globe. Al-Biruni wrote in total 150 works, including 35 treatises on pure astronomy, 
of which only six have survived. In addition, in the late $10^{\rm th}$ century Abu-al-Wafa and the prince Abu Nasr Mansur stated and proved 
theorems of plane and spherical geometry that could be applied by astronomers and geographers, including the laws of sines and tangents. 
Al-Biruni was Abu Nasr's pupil; he produced 
a vast amount of high-quality work, and was one of the masters in applying these theorems to astronomy and to such problems in mathematical geography 
as the determination of latitudes and longitudes, the distances between cities, and the direction from one city to another.

%
%\begin{figure}[hbt]
%\centering
%\includegraphics[width=4.5cm, height=2.8cm]{AlBiruni.jpg}
%\hskip 0.4cm
%\includegraphics[width=4.5cm, height=3cm]{al-biruni13.jpg}
%\caption{\small{\color{violet} {\bf Left: Al-Biruni; Right: Al-Biruni's commemorative stamp, Iran.}}}
%\label{fig:AlBiruni}
%\end{figure}
%
\vskip 0.1cm
\indent
$\bullet$ \; {\bf Al-Farghani} was one of Caliph Al-Mamun's astronomers. He wrote on the astrolabe, explaining the mathematical theory behind 
the instrument and correcting faulty geometrical constructions of the central disc, that were current then. His most famous book on cosmography,
{\it Kitab fi Harakat Al-Samawiyah wa Jaamai Ilm al-Nujum}, contains thirty chapters including a description of the inhabited part of the Earth, 
its size, the distances of the heavenly bodies from the Earth and their sizes, as well as other phenomena. 
\vskip 0.2cm
\indent
$\bullet$ \; {\bf Al-Zarqali} (Arzachel) ($1029-1087$) prepared the Toledan Tables and was also a renowned instrument maker 
who constructed a more sophisticated astrolabe: a {\it safiha}, accompanied by a treatise explaining how to use it. 
\vskip 0.2cm
\indent
$\bullet$ \; {\bf Jabir Ibn Aflah} (died in $1145$) was the first to design a portable celestial sphere to measure and explain the movements 
of celestial objects. Jabir is specially noted for his work on spherical trigonometry. Al-Bitruji's work ``{\it Kitab-al-Hay'ah}'' was translated 
by the Sicilian based Michael Scot, and bore considerable influence thereafter. 
\vskip 0.1cm
\indent
$\bullet$ \; {\bf Nur ad-Din al-Bitruji} (ca. $1150-1200$ C.E.), known in Europe as {\it Alpetragius}, is the author of {\it Kitab fi al-haya},
``A Book on Cosmology". He lived in Moorish Spain, in Andalusia, at the end of the Islamic Golden Age. He was probably a disciple of astronomer Ibn Tufayl. 
The problem faced by al-Bitruji was that faced by all Aristotelians who were reading Ptolemy's Almagest. 
Aristotle clearly stated that the planets must move with circular motions and implied that the center of these motions must be identical 
with the center of the Earth; he further desired a mechanism to transfer the motion of the prime mover to the planetary spheres. 

Ptolemy, on the other hand, while preserving the principle of circular motions (on eccentrics and epicycles), 
placed the centers of these motions elsewhere than at the center of the Earth; for Saturn, Jupiter, Mars, Venus, and Mercury 
he placed the centers of their uniform motions not at the centers of their respective eccentric deferents but at points called equants. 

Eudoxus of Cnidus had already shown that it is theoretically possible to explain the two most obvious anomalies in planetary 
motion: retrogression and latitude by means of homocentric spheres. Aristotle, by adding more spheres, converted this system 
to a mechanical model of the universe.
% (though technical details make it impossible for such a model to yield correct predictions 
%of the retrogressions and latitudes of Mars and Venus. 

Al-Bitruji followed the suggestion of Ibn Tufayl, as did the latter's other pupil Averroes, and attempted to adjust the Aristotelian solution 
in such a way that it would correspond to observed reality. 
The attempt failed owing to the inherent inadequacy of the homocentric system to describe the phenomena. 
\vskip 0.1cm
\indent
$\bullet$ \; {\bf Al-Tusi} ($1201 - 1274$), was the last great astronomer of Islamic Golden Age, and without exaggeration 
can be given the title of ``Hipparchus of Islamic Astronomy''. Of Persian origin, he studied in his native town Tus, then in the nearby town 
Nishapur, educated in  philosophy, medicine and mathematics. Already in Nishapur al-Tusi acquired a reputation 
as an oustanding scholar, and  became a member of Ismaili Court invited by the shi'ite ruler Abd ar-Rahman.  

%
%\begin{figure}[hbt]
%\centering
%\includegraphics[width=2.5cm, height=3.1cm]{Al-Tusi_portrait.jpg}
%\hskip 0.4cm
%\includegraphics[width=5.7cm, height=3.4cm]{Al_Tusi_stamp.jpg}
%\caption{\small{\color{violet} {\bf Left: the {\color{blue} Al-Tusi, a portrait},
% Right: al-Tusi commemorative stamp, IRAN.}}}
%\label{fig:AlTusi}
%\end{figure}

Al-Tusi stayed and worked in the Alamut castle until it fell to the troops of Dzhengis Khan's grandson Hulagu. Al-Tusi's reputation was so great 
that Hulagu named him his own advisor, and took him along to Baghdad, which was conquered by the Mongols in $1258$.  
Hulagu made Maragheh (in northwestern Iran) his capital, and the Observatory of the same name was built there. Under al-Tusi's direction, 
it has become a renowned center for mathematical and astronomical studies in the Islamic World. Al-Tusi designed many unique astronomical instruments, 
including a giant {\it azimuth quadrant} with radius of $4$ meters, and an improved astrolabe. 

Al-Tusi is mostly known for his geometric decomposition of linear motion into a sum of two circular motions, called ``the Tusi couple'', shattering
the established Aristotelean view on exclusivity of circular motions in heaven.
The same construction was used by Copernicus $250$ years later, probably re-invented independently. 
%\begin{figure}[hbt]
%\centering
%\includegraphics[width=3.5cm, height=3.2cm]{TusicoupleA.jpg}
%\hskip 0.2cm
%\includegraphics[width=3.6cm, height=3.5cm]{Tusicopernicus.jpg}
%\caption{\small {\color{violet} {\bf Left: {\color{red} the Tusi couple}, from his {\color{blue} {\it Memoir on Astronomy},}
% Right: the same construction from {\color{red} Copernicus' } {\color{blue} {\it De Revolutionibus}} {\color{black} $250$} years later.}}}
%\label{fig:absquarTusicouple}
%\end{figure}

Al-Tusi published the {\it Ilkhanic Tables}, relating his atsronomical observations made during $12$ years. Using his geometrical findings,
he improved substantially the model of lunar motion. He also determined the almost exact value of precession of the equinoxes, 
establishing it at $51'$ per century. {\it Commentary on the Almagest} contained excellent trigonometric tables
with values of chords calculated to three sexagesimal places for each half degree of argument. 

\vskip 0.1cm
\indent
$\bullet$ \; {\bf Ibn al-Shatir} ($1394-1375$) was a prominent astronomer and mathematician in Damascus, during his life under the rule
of the Ummayad dynasty. After completing studies in Cairo and Alexandria, he was appointed the {\it muwaqqit} at the great Ummayad mosque.
His most important book was an astronomical treatise entitled {\it ``The Final Quest concerning the Rectification of Principles'' }
contains new lunar, solar and planetary models, seriously departing from the Ptolemaic scheme, although firmly geocentric in principle. 

%
%\begin{figure}[hbt]
%\centering
%\includegraphics[width=4.8cm, height=3cm]{Ibn_Al-Shatir.jpg}
%\hskip 0.2cm
%\includegraphics[width=3.5cm, height=4.2cm]{Shatir_lunar.jpg}
%\caption{\small{\color{violet} {\bf  Left: Ibn al-Shatir, Right: A page from al-Shatir's book, with his lunar model.}}}
%\label{fig:AlShatir}
%\end{figure}
%
Ibn al-Shatir was not concerned with adhering to the theoretical principles of natural philosophy or Aristotelian cosmology, but rather to produce 
a model that was consistent with empirical observations. His concern for observational accuracy led him to eliminate the epicycles in the 
Ptolemaic solar model and all the eccentrics, epicycles and equant in the Ptolemaic lunar model. Shatir's new planetary model consisted 
of new secondary epicycles instead of equant, which improved on the Ptolemaic model.

\section{Epilogue}

Looking back at our sketchy summary of the history of Islamic science one can easily the major qualitative difference between its achievements
 in Mathematics versus all other domains. As a matter of fact, it is in Mathematics that the Arabs and Persians produced a genuine breakthrough
by adopting Indian system of notation and calculus, and by merging geometry and algebra together. This paved the way to the next revolutionary
breakthrough performed by the Europeans, who enlarged the scope of algebra y discovering complex numbers. This could not be achieved without
the algebraical basis led forth by the Islamic polymaths.  

On the other hand, with all its remarkable achievements, the Arab and Islamic Astronomy, especially in the observational field, represents the last chapter 
of the science of Antiquity, a natural extension of Babylonian; Indian, and above all Greek scientific heritages of a small world centered
on the Earth, for which the sphere of fixed stars represented the utmost external limit.

Modern science germinated within the new university framework of Medieval Europe between the $13^{\rm th}$ and $15^{\rm th}$ centuries,
openly born with the intellectual revolution brought by Copernicus in the $16^{\rm th}$ century, pursued later on by Galileo, Descartes,
Kepler, Leibniz, Newton and other European scientific luminaries.

\end{document}